\def\hgbr{$\kappa$-(BE\-DT\--TTF)$_2$\-Hg\-(SCN)$_{2}$\-Br}
\def\hgcl{$\kappa$-(BE\-DT\--TTF)$_2$\-Hg\-(SCN)$_{2}$\-Cl}
\def\cubr{$\kappa$-(BE\-DT\--TTF)$_2$\-Cu\-[N\-(CN)$_{2}$]\-Br}
\def\cucl{$\kappa$-(BE\-DT\--TTF)$_2$\-Cu\-[N\-(CN)$_{2}$]\-Cl}

\def\cucn{$\kappa$-(BE\-DT\--TTF)$_2$\-Cu$_2$(CN)$_{3}$}
\def\agcn{$\kappa$-(BE\-DT\--TTF)$_2$\-Ag$_2$(CN)$_{3}$}

\def\cm{cm$^{-1}$}
\documentclass[aps,prb,superscriptaddress,twocolumn,showpacs]{revtex4-1}
\usepackage{graphicx}%

\begin{document}

\title{Metal-Insulator Transition in the Dimerized Organic Conductor $\kappa$-(BEDT-TTF)$_2$Hg(SCN)$_2$Br}
\author{Tomislav Ivek}
\affiliation{1.\ Physikalisches Institut, Universit{\"a}t Stuttgart, Pfaffenwaldring 57, 70550 Stuttgart, Germany}
\affiliation{Institut za fiziku - P.O.\ Box 304, HR-10001 Zagreb, Croatia}
\author{Rebecca Beyer}
\affiliation{1.\ Physikalisches Institut, Universit{\"a}t Stuttgart, Pfaffenwaldring 57, 70550 Stuttgart, Germany}
\author{Sabuhi Badalov}
\affiliation{1.\ Physikalisches Institut, Universit{\"a}t Stuttgart, Pfaffenwaldring 57, 70550 Stuttgart, Germany}
\author{Matija \v{C}ulo}
\affiliation{Institut za fiziku - P.O.\ Box 304, HR-10001 Zagreb, Croatia}
\author{Silvia Tomi{\'c}}
\affiliation{Institut za fiziku - P.O.\ Box 304, HR-10001 Zagreb, Croatia}
\author{John A. Schlueter}
\affiliation{Division of Materials Research,
National Science Foundation, Arlington, VA 22230, and
Material Science Division, Argonne National Laboratory,
Argonne, Illinois 60439-4831, U.S.A.}
\author{Elena I. Zhilyaeva}
\affiliation{Institute of Problems of Chemical Physics, Russian Academy of Sciences, RU-142~432 Chernogolovka, Moscow oblast, Russia}
\author{Rimma N. Lyubovskaya}
\affiliation{Institute of Problems of Chemical Physics, Russian Academy of Sciences, RU-142~432 Chernogolovka, Moscow oblast, Russia}
\author{Martin Dressel}
\affiliation{1.\ Physikalisches Institut, Universit{\"a}t Stuttgart, Pfaffenwaldring 57, 70550 Stuttgart, Germany}
\date{\today}

\begin{abstract}
The organic charge-transfer salt $\kappa$-(BEDT-TTF)$_{2}$Hg(SCN)$_{2}$Br
is a quasi two-dimensional metal with a half-filled conduction band at ambient conditions.
When cooled below $T=80$\,K it undergoes a pronounced transition to an insulating phase
where the resistivity increases many orders of magnitude.
In order to elucidate the nature of this metal-insulator transition
we have performed comprehensive transport, dielectric and optical investigations.
The findings are compared with
other dimerized $\kappa$-(BEDT-TTF) salts, in particular the Cl-analogue, where a charge-order
transition takes place at $T_{\rm CO}=30$\,K.
\end{abstract}

\maketitle

\section{Introduction}
The current interest in dimerized two-dimensional organic conductors mainly focusses on
$\kappa$-(BEDT-TTF)$_2X$ salts built with extended polymeric anions $X^{-}$ such as
Cu(SCN)$_2^-$, Cu[N(CN)$_2$]Br$^-$, Cu[N(CN)$_2$]Cl$^-$, or Cu$_2$(CN)$_3^-$,
all containing copper ions.
These systems are characterized by half-filled conduction bands and are located in proximity to correlated insulating states. Depending on the strength of the on-site Coulomb repulsion $U$ with respect to the bandwidth $W$, a metallic and superconducting ground state develops at low temperatures, or a Mott insulator, which might be antiferromagnetically ordered or
behaves like a spin liquid due to strong frustrations.\cite{Ishiguro98,Kanoda11,Powell11}

Subsequent theoretical approaches \cite{Hotta10,Naka10,Gomi10,Dayal11}
suggest that also the Coulomb interaction $V$ between the dimers has to be included in the description of possible charge-ordering phenomena in these dimerized (BEDT-TTF) salts.
Although numerous experimental results,\cite{Sedlmeier12,Tomic13,Pinteric14,Shimizu06,Yakushi15,Dressel16} rule out appreciable charge disproportionation in \cucl, \cucn{} and other copper-based $\kappa$-(BEDT-TTF)$_2X$ salts, these considerations might be relevant for dimerized salts in general.
Unfortunately, charge-order phenomena in $\kappa$-phase salts with an
effectively half-filled band are scarce and only a few compounds are
reported exhibiting charge order, such as
$\kappa$-(BEDT\--TTF)$_4$\-PtCl$_6$$\cdot$C$_6$H$_5$CN, the
triclinic
$\kappa$-(BEDT\--TTF)$_4$\-[$M$(CN)$_6$]\-[N(C$_2$H$_5$)$_4$]$\cdot$3H$_2$O
and the monoclinic
$\kappa$-(BEDT\--TTF)$_4$[$M$(CN)$_6$]\-[N(C$_2$H$_5$)$_4$]\-$\cdot$2H$_2$O
(with $M$ =  Co$^{\rm III}$, Fe$^{\rm III}$, and Cr$^{\rm III}$)
salts.\cite{Doublet94} Here the structure is rather complex as the phase
transition includes the deformation of the molecule and the
interaction with the anions; accordingly, details of their physical
properties and their electronic states are not well known.
Certainly electronic correlations as well as coupling to the
lattice are important.

\begin{figure}
    \centering
        \includegraphics[width=0.7\columnwidth]{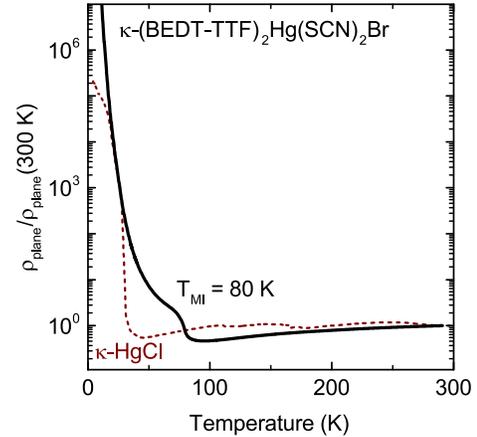}
    \caption{Temperature dependence of the normalized dc resistivity of \hgbr{}
    measured within the highly conducting $(bc)$ plane.
    By cooling below $T_{\rm MI}=80$\,K a metal to insulator transition occurs.
    For comparison the normalized dc resistivity of sibling compound \hgcl{} is plotted as dashed line;
    due to charge ordering the system becomes insulating below $T_{\rm CO} = 30$\,K.}
    \label{fig:dc}
\end{figure}
Recently it was demonstrated that \hgcl{} undergoes
a pronounced charge order transition at $T_{\rm CO}=30$\,K,\cite{Yasin12,Drichko14}
as illustrated in Fig.\ \ref{fig:dc}.
The insulating state is rapidly suppressed by hydrostatic pressure of $p=4.2$\,kbar without any signs of superconductivity;\cite{Lohle17} making this compound distinct from \cucl, where a tiny amount of pressure drives the Mott insulator superconducting, but also from the canonical charge-ordered system $\alpha$-(BEDT-TTF)$_2$I$_3$.\cite{Dressel94,Beyer16}
The present investigation now addresses the question in which way $\kappa$-(BEDT-TTF)$_2$Hg(SCN)$_2$$X$ is modified
when replacing $X$ = Cl by Br in the anion sheet. Let us recall that in the case of \cucl{} the corresponding substitution
pushes the antiferromagnetic Mott insulator into a superconducting phase with $T_c \approx 12$\,K.\cite{Ishiguro98}

The family of $\kappa$-(BEDT-TTF)$_2X$ salts with mercury-based anions have a
structure similar to the Cu-salts,\cite{Lyubovskii96} but different ratios of transfer integrals, which result in modified parameters used to map them on the Hubbard
model.\cite{Jeschke}
In particular the orbital overlap $t_d$ within the dimers of the Hg-compounds is weaker compared to the the Cu-family,
causing a reduced  on-site repulsion $U$ with respect to the inter-dimer interaction.
This makes it necessary to go beyond  the
simple Hubbard model, which allowed to treat the Cu-based $\kappa$-(BEDT-TTF) compounds
successfully as half-filled system with one electron per dimer.\cite{Seo04,Powell11}
Now the description has to start from individual BEDT-TTF molecules, leading to a quarter-filled conduction band; in addition to $U$ the inter-molecular interaction $V$ has to be included in the extended Hubbard model.

In an effective dipolar-spin model Hotta suggests \cite{Hotta10} that
quantum electric dipoles are formed on the dimers, which interact with each other
and thus modify the exchange coupling.
Since they fluctuate by $t_d$, for large orbital overlap the dimer Mott insulator is stable, forming a dipolar liquid; if $V$ is large compared to $t_d$, however, charge
order emerges (dipolar solid). Similar considerations have been
put forward by other groups.\cite{Naka10,Gomi10}
Mazumdar, Clay and collaborators \cite{Dayal11} could show
that in these system a frustration-induced transition occurs from a N{\'e}el
antiferromagnetism to a spin-singlet state in the interacting
quarter-filled band on an anisotropic triangular lattice. In the
spin-singlet state the charge on the molecules becomes unequal: a
paired-electron crystal is formed with pairs of charge-rich sites
separated by pairs of charge-poor sites. This spin-singlet
formation is driven by quantum effects at any value of nearest
neighbor interaction $V$.

Here we present the first comprehensive transport, dielectric and optical investigations of \hgbr\
and discuss the findings within the framework of dimerized Mott insulators
composed by these Hg- and Cu-families.

\section{Experimental Details}
A  series of $\kappa$-(BE\-DT\--TTF)$_2$\-Hg(SCN)$_{3-n}X_n$ salts
with $X$ = F, Br, I ($n = 1$) and $X$ = Cl ($n = 1, 2$)
has been synthesized by Lyubovskii and collaborators
already twenty years ago,\cite{Aldoshina93,Konovalikhin93}
ranging from insulators to metals and possible superconductors;\cite{Lyubovskii96}
a thorough investigation of their physical properties, however, is still lacking.
Here we have turned our attention to \hgbr, an isostructural sister compound of
the charge-ordering \hgcl; as demonstrated in Fig.\ \ref{fig:dc}, it also exhibits
a metal-insulator transition upon cooling, but at somewhat higher temperatures.

\begin{figure}
    \centering
        \includegraphics[width=0.9\columnwidth]{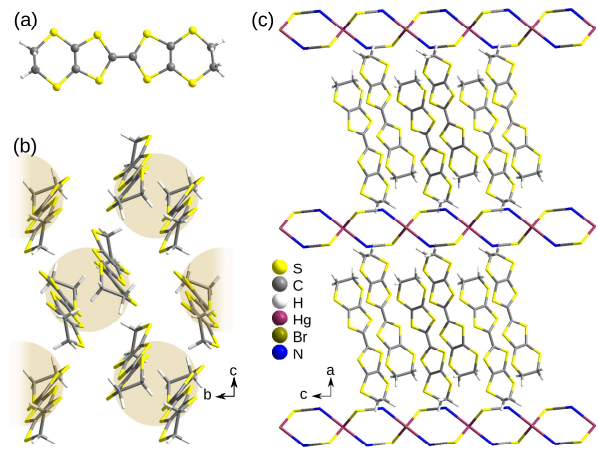}
    \caption{(Color online) (a) Sketch of the bis-(ethyl\-ene\-di\-thio)\-te\-tra\-thia\-ful\-va\-lene
molecule, called BEDT-TTF.
(b) The view on the BEDT-TTF layers along the out-of-plane $a$ direction illustrates the dimer structure (dimers are denoted by shaded circles).
(c) The molecular layers along the conducting $bc$ plane are separated by the Hg(SCN)$_2$Br$^-$ anions. Note that within a dimer
the two molecules are displaced along the molecular axis, leading to a reduced transfer
integral $t_d$.}
    \label{fig:structure}
\end{figure}

Based an recent x-ray scattering experiments at ambient conditions,
the crystal structure of \hgbr{} is depicted in Fig.\ \ref{fig:structure}
along two crystallographic directions.
The BEDT-TTF molecules are all crystallographically equivalent and within the $bc$-plane form dimers according to the $\kappa$-pattern\cite{Mori99} that are rotated with respect to each other.
As seen from Fig.\ \ref{fig:structure}(b) the stacking direction $b$ is more pronounced
than known from other $\kappa$-salts
leading to a rather isotropic response in our transport and optical measurements.

It should be noted that during the electrochemical synthesis, single crystals of the $\beta^{\prime\prime}$-phase tend to grow as well to a considerable size (1--2\,mm),\cite{Li16} and subsequently have to be separated from the desired $\kappa$-phase specimens with the help of x-ray or infrared optical inspection.

In the temperature-dependent dc resistivity  of \hgbr{} displayed in Fig.\ \ref{fig:dc}
a sharp metal-insulator transition is identified at $T=80$\,K, confirming the early characterization.\cite{Aldoshina93} While for \hgcl{} a saturation of $\rho(T)$ is observed at very low temperatures, for the Br-analogue the slope $|{\rm d}\rho(T)/{\rm d}T|$ continues to rise as the temperature is lowered.
For a more complete study we have measured the charge transport within the highly-conducting $(bc)$ plane, as well as perpendicular to it, using the standard four-probe technique. The sample was slowly cooled (up to 0.5\,K/h) from room temperature down to $T=4$\,K using a home-made helium bath cryostat.

In addition, the spectra of the complex dielectric function were obtained from measuring the two-contact complex conductance in the frequency range from 20\,Hz to 10\,MHz for various temperatures. The Hewlett-Packard 4284A LCR meter and Agilent 4294A impedance analyzer with virtual ground method were used. The capacitive open-loop contribution of the sample holder was always subtracted.

The optical properties of \hgbr{} have been investigated by infrared reflectivity measurements from room temperature down to $T=4$\,K for both polarizations along the main axes of the highly-conducting $(bc)$ plane.
To this end a Bruker Hyperion infrared microscope was attached to the Fourier-transform spectrometer Bruker IFS 66v/s or Vertex 80v.
The sample was cooled down to helium temperatures by a Cryovac Microstat cold-finger cryostat.
The far-infrared data were taken by a Bruker IFS 113v equipped with a
cold-finger cyostat and in-situ gold evaporation as reference.\cite{Dressel04}
In order to perform the Kramers-Kronig analysis,
a constant reflectivity extra\-po\-lation was used at low frequencies and
temperatures $T_{\rm MI}<80$\,K, while a Hagen-Rubens behavior was assumed for elevated
temperatures.\cite{DresselGruner02}
In addition, the vibrational features were measured perpendicular to the plane ($E\parallel a$) using an infrared microscope.
In particular the {\em ungerade} C=C vibration $\nu_{27}({\rm b}_{1u}$) of the BEDT-TTF molecules are utilized as a local probe to determine the charge per molecule \cite{Dressel04,Drichko09,Yamamoto05,Girlando11} and follow any possible charge disproportionation.

\section{Results and Analysis}

\subsection{Transport Properties}
\begin{figure}
    \centering
        \includegraphics[width=0.8\columnwidth]{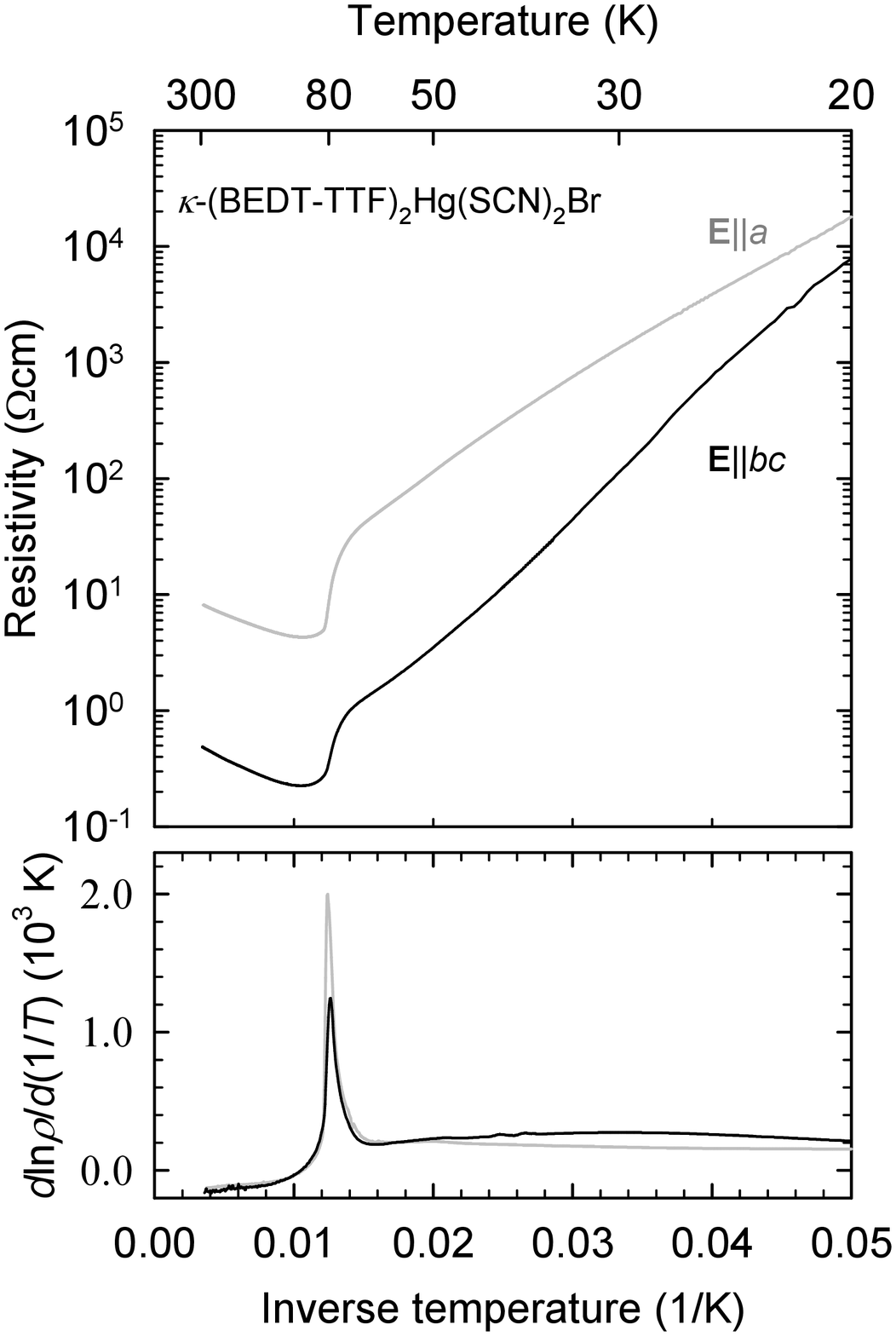}
    \caption{(a) Electrical resistivity of \hgbr{} as a function of inverse temperature measured within the $(bc)$-plane and perpendicular to the plane, i.e., $E\parallel a$. (b) The phase transition around $T_{\rm MI} \approx 80$\,K is identified as the maximum in the derivative.}
    \label{fig:HgBr-dc}
\end{figure}
In order to gain more information on the transport mechanisms of \hgbr{} above and below the phase transition at $T_{\rm MI} = 80$\,K, we have performed dc resistivity measurements along different crystal directions and present the results  down to $T=20$\,K in an Arrhenius plot  in Fig.\ \ref{fig:HgBr-dc}(a) as a function of inverse temperature. At ambient conditions, the resistivity within the $(bc)$-plane is approximately 0.5\,$\Omega$cm with an anisotropy of less than a factor of 2. Perpendicular to the highly-conducting plane, i.e., parallel to the $a$-axis,
$\rho_{\rm dc}$ is about 20 times higher.
For all orientations a very similar temperature dependent metallic resistivity is observed down to 100\,K.
At $T_{\rm MI}$ a metal-insulator transition occurs in all three directions with a more or less steep rise in $\rho(T)$ by an order of magnitude or more. It is interesting to note that this temperature-behavior very much resembles the metal-insulator transition observed in other ET-based compounds such as $\alpha$-(BEDT-TTF)$_2$I$_3$ and $\theta$-(BEDT-TTF)$_2$RbZn(SCN)$_4$,\cite{Tomic15} where the gap at the density of states rapidly opens at the transition.
The insulating state is characterized by an activated behavior $\rho(T)\propto \exp\{E_g/2k_BT\}$ down to lowest temperatures; no saturation is observed up to our limit of $10^9\,\Omega$cm.
Within the plane ($E\parallel bc$) the activation energy of $E_g/2=(22\pm 3)$\,meV is basically constant down to $T= 20$\,K. In the out-of-plane direction ($E\parallel a$), we extract a similar activation energy of $E_g/2=(15.5\pm 2)$\,meV, which means it is rather isotropic. If we consider $E_g$ as the full gap in the density of states, the ratio $E_g/k_B T_{\rm MI}\approx 4-5.6$ is not much higher than 3.53 expected from a conventional mean-field transition. Thus we can consider this system mean-field-like with relatively weak electronic correlations.

\subsection{Dielectric Properties}
\begin{figure}
    \centering
        \includegraphics[width=0.8\columnwidth]{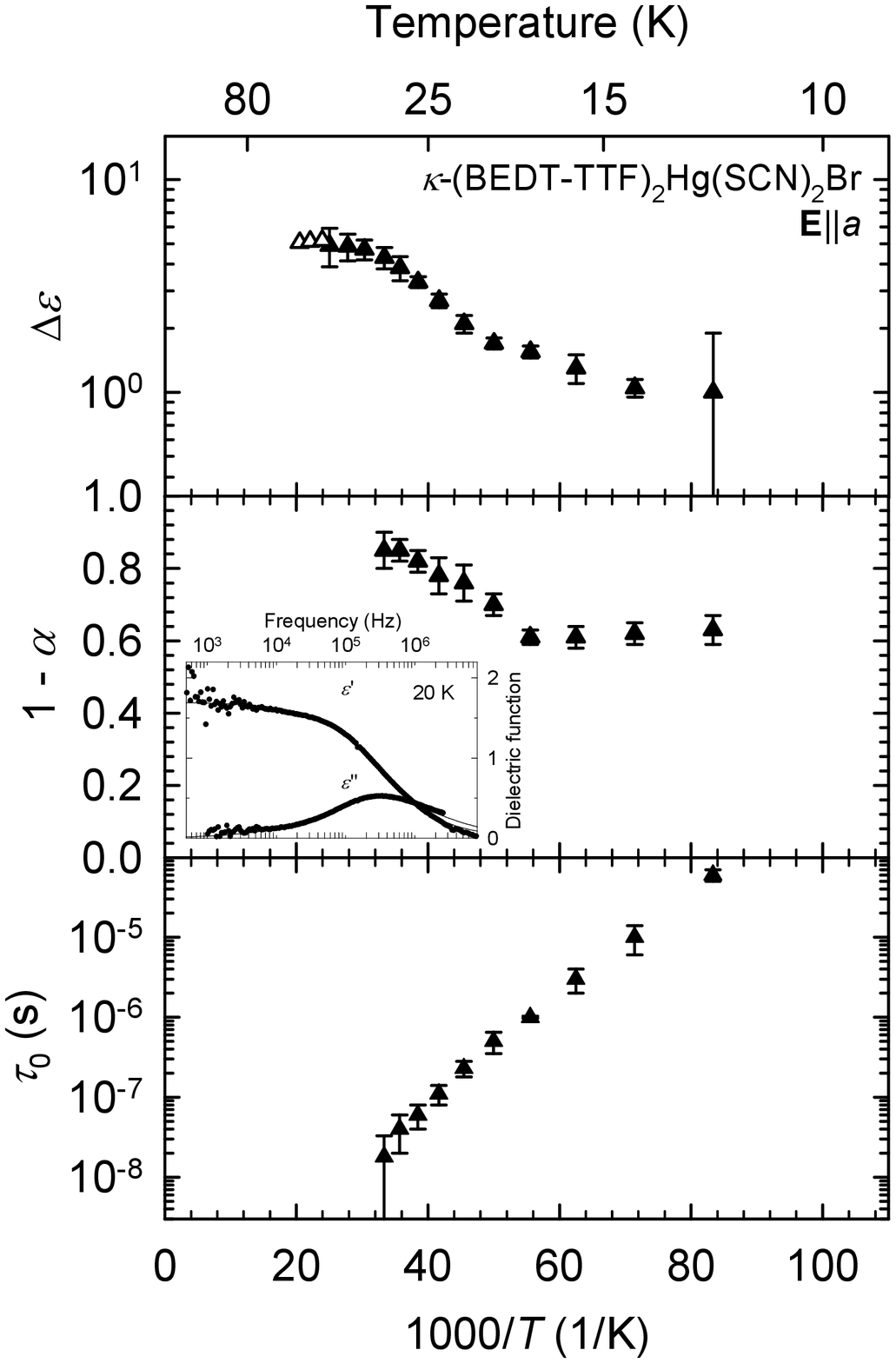}
    \caption{(a) Delectric strength $\Delta\varepsilon$, (b) distribution width of relaxation times $1-\alpha$, and (c) mean relaxation time $\tau_0$ of \hgbr{} as a function of inverse temperature measured perpendicular to the $(bc)$-plane. Inset: real and imaginary part of the measured dielectric response (points) and fit (lines, see text) at a select temperature.}
    \label{fig:ds_param}
\end{figure}
Like in many other charge-ordered (BEDT-TTF) compounds,\cite{Ivek11,Tomic13,Pinteric14,Pinteric16} the dielectric properties of \hgbr{} can be described by a generalized Debye expression
\begin{equation}
\varepsilon(\omega) - \varepsilon_{\rm HF} =
\frac{\Delta\varepsilon}{1 + \left({\rm i}\omega \tau_0\right)^{1-\alpha}}
\end{equation}
at low temperatures, where $\Delta\varepsilon  = \varepsilon_0 - \varepsilon_{\rm HF}$.
corresponds to the strength of the dielectric mode;  $\varepsilon_0$ and $\varepsilon_{\rm HF}$ are the static and high-frequency dielectric
constants, respectively; $\tau_0$ is the mean relaxation time;
and $1-\alpha$ is the symmetric broadening of the relaxation time distribution.
The temperature dependences of the extracted parameters $\Delta\varepsilon$, $1-\alpha$,
and $\tau_0$ are plotted in Fig.\ \ref{fig:ds_param} as a function of inverse temperature $1/T$. The dielectric relaxation can be detected only at temperatures below 60\,K.
Above $T\approx 35$\,K we can determine only the dielectric relaxation strength
by measuring the capacitance at 1\,MHz.
At first glance the relaxation appears rather broad. The dielectric strength perpendicular to planes ($E \parallel  a$) is only of the order of 10 and less while the strength of the dielectric response $E \parallel bc$ is on the order of 100 (not shown), which mirrors the similarly weak anisotropy found in dc data.
Most important, for both orientations the intensity of the modes becomes smaller by about an order of magnitude as the temperature is reduced to 10\,K; these findings are in contrast to the behavior commonly observed in low-dimensional organic charge transfer salts.\cite{Tomic15}
It clearly indicates that \hgbr{} does not exhibit indications of charge order and shows only very weak relaxor-like properties, further underlining the absence of strong electronic correlations.

For both orientations, the temperature behavior of $\tau_0(T)$ shows no saturation and basically follows an Arrhenius law between 40 and 12\,K, as plotted in Fig.\ \ref{fig:ds_param}(c). The activation energy is comparable with the dc resistivity. This indicates that electronic screening by quasi-free charge carriers is the dominant relaxation mechanism.

\begin{figure}
    \centering
        \includegraphics[width=1.00\columnwidth]{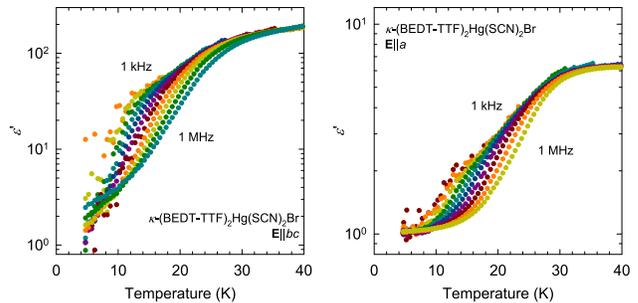}
    \caption{(Color online) Real part of the dielectric constant $\varepsilon(T)$ of \hgbr{} determined (left panel) within the $(bc)$-plane and (right panel) for $E\parallel a$ (perpendicular to the plane) at various frequencies from 1\,kHz up to 1\,MHz as indicated. Note the logarithmic scale.}
    \label{fig:ds_Tsweep}
\end{figure}
Around $T\approx 20 - 30$\,K we find the most pronounced change in the dielectric behavior that can be seen even better by directly looking at the temperature dependence of the dielectric constant. In Fig.\ \ref{fig:ds_Tsweep}, we present $\varepsilon'(T)$ measured at different frequencies for the orientations parallel and perpendicular to the $(bc)$-plane. A clear step can be seen around $T=20$\,K that becomes more pronounced as the frequency increases to 1\,MHz. This is accompanied by a change in $1-\alpha$ and $\Delta\varepsilon$ at the same temperatures in Fig.\ \ref{fig:ds_param}, which suggests the entities responsible for the dielectric response are not correlated as temperature is lowered.

\subsection{Infrared-active Molecular Vibrations}
Microscopic information on the charge distribution and its temperature behavior
can be obtained by optical spectroscopy.
To this end we performed infrared reflectance measurements
for the polarization perpendicular to the layers,
i.e., $E\parallel a$.
In this orientation the optical response resembles an insulator
with low reflectivity and correspondingly low values of conductivity,
in agreement with the quasi-two-dimensional character of these materials,
deduced from the dc conductivity presented in Fig.\ \ref{fig:HgBr-dc}.

Of primary interest, however, are the superimposed infrared-active
vibrational features of the BEDT-TTF molecule and lattice.
The out-of-phase molecular vibrations of the two C=C bonds of the inner rings
provide a powerful and well-established method to identify the amount of charge
on molecular lattice sites.\cite{Dressel04,Drichko09,Girlando11}
The frequency dependence of the $\nu_{27}({\rm b}_{1u})$ mode
allows for a quantitative determination of the charge per BEDT-TTF molecule
and thus the most accurate way to monitor the temperature-dependence of
the charge disproportionation.
\begin{figure}
    \centering
        \includegraphics[width=0.8\columnwidth]{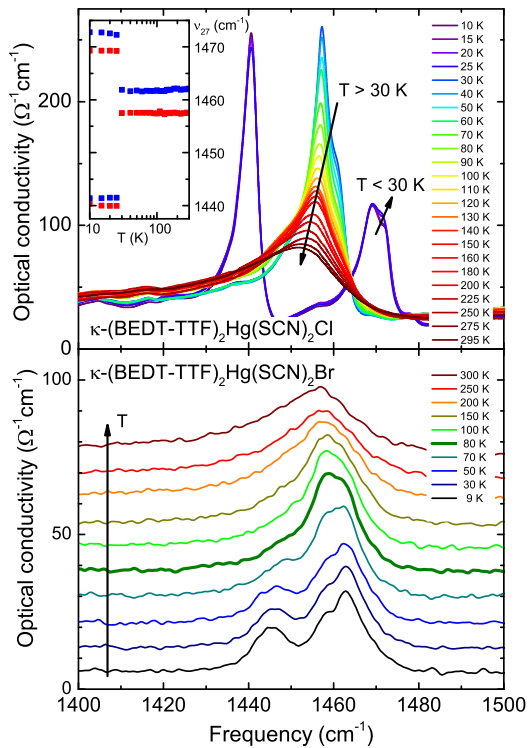}
    \caption{(Color online) Temperature dependence of the absorption feature related to the molecular vibration $\nu_{27}({\rm b}_{1u})$ which is seen as a local probe of the charge per BEDT-TTF molecule. In the upper panel a clear splitting is observed in  \hgcl{} as the temperature is reduced below the charge order transition $T_{CO}=34$\,K. The inset displays the peak frequency as a function of temperature. The lower panel shows the $\nu_{27}$ vibrational mode of \hgbr{} for different temperautre in a waterfall plot; i.e., the curves are shifted with respect to each other. Except of a gradual development of a sidepeak, no splitting due to charge order is found.}
    \label{fig:nu27}
\end{figure}

The resulting conductivity spectra in the region of
the $\nu_{27}({\rm b}_{1u})$ mode are presented in Fig.\ \ref{fig:nu27} for
\hgbr{} in comparison with spectra of the Cl-analogue.
\hgcl{} is known \cite{Drichko14} to undergo a charge-order transition around  $T_{\rm CO}=30$\,K, as can be clearly seen from the splitting of the $\nu_{27}({\rm b}_{1u})$ vibration, which
peaks around  1455\,\cm{} at room temperature and narrows on cooling down.
A higher-frequency shoulder can be identified that becomes more pronounced at low temperatures; this double features indicates the two crystallographically different sites
per unit cell. For a quantitative analysis we fit the bands by two Fano resonances
for $T>30$\,K and four modes below $T_{\rm CO}$; the results are plotted in the inset
and perfectly agree with our previous observations \cite{Drichko14}
where we concluded a charge difference of $2\delta_{\rho} = 0.2e$ between
two different molecular sites. L{\"o}hle {\it et al.} studied the pressure dependence of the charge ordered phase \cite{Lohle17} and found a rapid suppression of $T_{\rm CO}$ by about 0.7\,kbar and the absence of charge disproportionation for $p>4$\,kbar.

For the present case of \hgbr{} nothing like that is observed in the lower panel of Fig.\ \ref{fig:nu27}. At room temperature the
broad $\nu_{27}({\rm b}_{1u})$ feature occurs at 1457\,\cm with a comparable width and temperature dependence down to $T\approx 100$\,K. No drastic change is observed when the sample is cooled further below $T_{\rm MI}$, in a way similar to Cu-based $\kappa$-BEDT-TTF$_2X$ systems.\cite{Sedlmeier12} At the lowest temperature, the double peak structure around 1460\,\cm{} has slightly changed in so far as the higher-frequency peak is more pronounced. In addition, a second feature develops at 1445\,\cm at below temperatures as high as 150\,K, meaning it is not related to the metal-insulator transition. Even though it is tempting to ascribe its temperature evolution to the establishing of the low-temperature phase, it appears to be governed by thermal broadening: the width and strength behave in a manner similar to the dominant 1460\,\cm{} peak as well as the normal-phase $\nu_{27}({\rm b}_{1u})$ peak of the \hgcl{} analogue.

It is interesting to note that the $\nu_{27}$ mode observed in \hgbr{} is much broader than in the Cl-analogue and comparable to the ones found in the spin-liquid candidates \cucn{} and \agcn{},\cite{Sedlmeier12,Pinteric16} where charge inhomogeneities or fluctuations might be of relevance.\cite{Tomic15,Pinteric16} These findings provide evidence that the metal-insulator transition observed in \hgbr{} at 80\,K is not due to static charge ordering. Recent measurements of the Raman spectra as a function of temperature support this conclusion: no indications of charge imbalance has been observed \cite{Hasan17} as the metal-insulator transition is passed.

\subsection{In-Plane Infrared Properties}
\begin{figure*}
    \centering
        \includegraphics[width=0.8\textwidth]{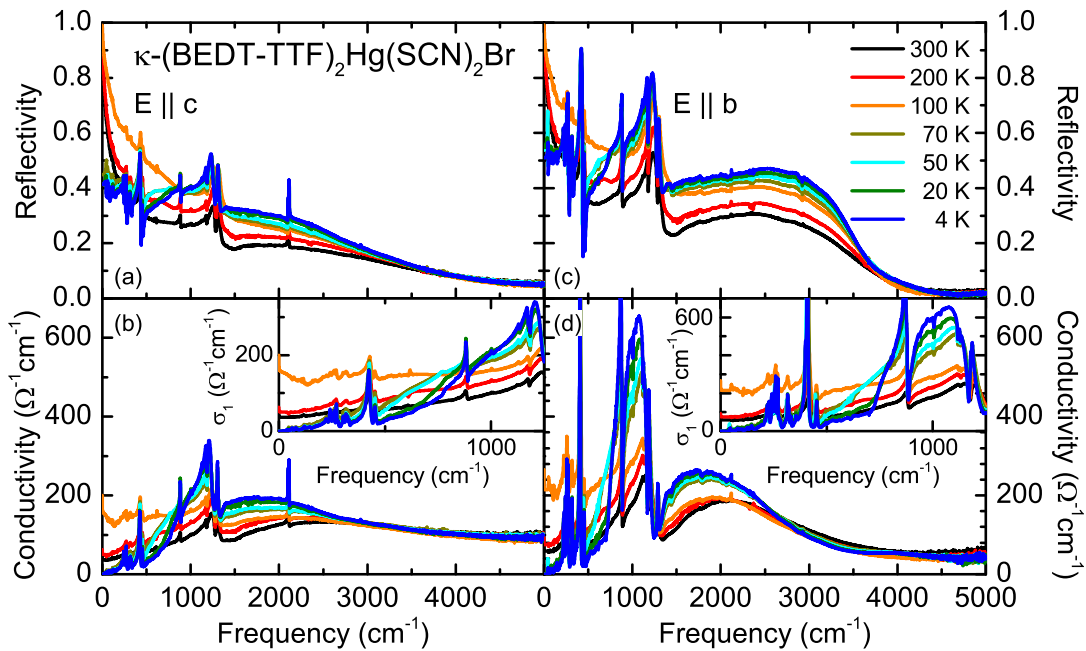}
    \caption{(Color online) Optical reflectivity of \hgbr{} measured at different temperatures as indicated
    (a) for light polarized along the $c$ direction and
    (c) for $E\parallel b$. (b,d) The corresponding optical conductivity as obtained by
    Kramers-Kronig analysis for the two different orientations. The insets show the low-frequency conductivity $\sigma_1(\omega)$ on an enlarged scale.}
    \label{fig:RefCond}
\end{figure*}
The electrodynamic properties of \hgbr{} were investigated by measuring the reflectivity at different temperatures for  $4\,{\rm K} \leq T < 300$\,K;
in Fig.\ \ref{fig:RefCond} we plot $R(\omega)$  for the two polarizations $E\parallel c$ and $E\parallel b$ together with the optical conductivity obtained by the Kramers-Kronig relations. The quasi-two-dimensional conductor displays a rather weak anisotropy in its optical response, very similar to its Cl-analogue \hgcl{} and to $\kappa$-(BEDT-TTF) salts with Cu-containing anions. From the crystal structure [Fig.\ \ref{fig:structure}(b)] we can identify the $b$-axis as preferred stacking direction of
the BEDT-TTF molecules.

Most obvious, the mid-infrared band is more intense along the $b$-direction [Fig.\ \ref{fig:RefCond}(c,d)], leading to a drop in $R(\omega)$ around 3500\,\cm{} that resembles the plasma edge of regular metals. We associate this optical anisotropy to the arrangement of the BEDT-TTF dimers as depicted in Fig.\ \ref{fig:structure}(b): the intra-dimer charge transfer is in particular excited by $E\parallel b$.
The low-frequency reflectivity is rather small, between 0.4 and 0.6. Only when cooled down to $T=100$\,K a pronounced Hagen-Rubens behavior is measured. This corresponds to the increase in conductivity by a factor of 3 to 4 (see insets of Fig.\ \ref{fig:RefCond}), in good agreement with the metallic temperature dependence seen in dc resistivity (Figs.\ \ref{fig:dc} and \ref{fig:HgBr-dc}) above the metal-to-insulator transition.

From the mid-infrared conduction band we can extract the electronic parameters,\cite{Dressel04,Faltermeier07,Merino08,Dumm09} in particular the maximum $\nu_{\rm max}$ is proportional to the Coulomb repulsion $U$. At $T=100$\,K we find approximately
2200\,\cm{} for $E\parallel c$, while for $E\parallel b$ it is only 1950\,\cm. In both cases the values increase by about 10\%{} as the temperature rises to room temperature. The overall behavior is similar to the observations reported for \hgcl, however, the absolute values are reduced, indicating smaller effect of electronic correlations.

We describe the zero-frequency contributions by a Drude-like conductivity. In accord to the dc data, the spectral weight $\int\sigma(\omega){\rm d}\omega = \omega_p/8$ obtained for both directions differs by a factor of 2, independent on temperature. For most-conducting direction ($E\parallel b$) the plasma frequency $\omega_p/(2\pi c) = 1500$\,\cm{} at ambient temperature increases to 4000\,\cm{} at $T=100$\,K. Here the width increases by a slightly more than a factor of 2 when the temperature is reduced to $T_{\rm MI}$. Within the uncertainty, no change is observed along the $c$-direction.

For \hgbr{} we observe a larger spectral weight of the Drude component compared to the one extracted for the Cl-analogue, corroborating our conclusion drawn above from the mid-infrared band, that the title compound is less-correlated than \hgcl. A similar order was observed for the copper-based salts, as \cucl{} is a Mott insulator while the Br-analogue becomes superconducting.

For $T<T_{\rm MI}$ the low-frequency reflectivity drops significantly in both polarizations. Although strong phonon bands around 200 - 450\,\cm{} cause considerable intensity in $\sigma_1(\omega)$ even at $T=4$\,K,  we might identify an isotropic gap around 500\,\cm{} for both directions. Within experimental errors, this value agrees nicely with the transport gap extracted from dc data. Additionally, we note that frequencies up to 1000\,\cm{} the optical conductivity decreases when cooling from 70 to 4\,K which indicates a sort of soft gap behavior.
Accordingly the spectral weight is shifted to the mid-infrared band, which grow considerably.

\section{Discussion}
By now it is well established that \hgcl{} undergoes a metal-insulator transition due to charge ordering at $T_{\rm CO}=30$\,K: the resistivity $\rho_{\rm dc}(T)$ continuously rises by orders of magnitude (Fig.\ \ref{fig:dc}) and the charge-sensitive vibrational $\nu_{27}({\rm b}_{1u})$-mode splits (Fig.\ \ref{fig:nu27}) indicating a charge disproportionation $2\delta_{\rho}=0.2e$ at low temperatures.
In contrast, no clear splitting of the $\nu_{27}({\rm b}_{1u})$-mode is observed in the sibling \hgbr, ruling out any significant charge disproportionation. Hence, \hgbr{} does not enter a charge-ordered ground state.

Applying minor pressure, the Mott insulator with Cu-based anions \cucl{} turns metallic and superconducting, very much like \cubr{} at ambient pressure. In the case of the Hg-based \hgcl{} the relation to the Br-analogue is rather different: pressure rapidly suppresses the charge-ordered state in \hgcl{}, which becomes metallic at all temperatures.\cite{Lohle17} This is in striking contrast to \hgbr{} which shows no charge ordering but still features a metal-insulator transition of a different kind.
 
It has been suggested \cite{Hotta10} that charge fluctuations within the dimers may couple to neighboring entities; and in particular in the frustrated case of a triangular lattice it may give rise to the spin liquid state. Pressure enhances the interdimer interaction and may explain why \hgcl{} exhibits a strong and static charge order, while only charge fluctuations may be present in \hgbr.

For the organic Mott insulators \cucl{} and \cucn{} the temperature evolution of dielectric spectra resembles relaxor ferroelectrics.\cite{Abdel-Jawad2010,Lunkenheimer12,Tomic13,Pinteric14}
In the case of \hgcl{} the behavior is similar to other charge-ordered compounds, such as $\alpha$-(BEDT-TTF)$_2$I$_3$,\cite{Ivek11,Ivek17} where two dielectric relaxation modes appeared
in the kHz-MHz frequency range. Nothing like that however is observed in \hgbr{}, supporting our conclusion that the compound is not a charge-order insulator, but also that it is not clear whether the disorder in the molecular and anion layers is an issue here. The temperature-dependence of the dielectric parameters indicate some relaxational behavior screened by the conduction electrons remaining below the phase transition. It is 30 times more pronounced within the $(bc)$-plane compared to the perpendicular direction. Below 30\,K the mode vanishes rapidly and is basically absent at $T=10$\,K, although in this temperature range no additional transition can be identified in the transport or optical properties. It might be tempting to attribute the dielectric response to dipolar fluctuations,\cite{Tomic15} forming a kind of non-Barrett quantum electric-dipole liquid.\cite{Shen16} Since pure dipolar coupling does not cause frustration in a two-dimensional triangular lattice, the transfer integrals are significant and do increase when going from \hgbr{} to the Cl-analogue.

The low-temperature optical conductivity of \hgbr{} exhibits a shape similar to
the room-temperature spectra of other $\kappa$-(BEDT-TTF) salts, such as \cucl,
with a gap at low frequencies, strong electron-molecular vibrational (emv) coupled features, and a band at about 2000\,\cm. The electronic properties of these compounds can be well described by assuming a lattice of dimers with one electron per site. The simple Hubbard model of a half-filled system describes the main physics rather well.
In comparison to \hgcl{} and \cucl{}, the effect of electronic correlations seems to be less important in \hgbr. This is concluded from the mean-field-like dc transport and maximum of the mid-infrared conductivity band, but also from the spectral weight of the Drude component (Fig.\ \ref{fig:RefCond}). The zero-frequency peak is higher than typically observed in the Cu-family and it shows up at high temperatures, as a result of smaller Hubbard on-site repulsion $U$.\cite{Merino08,Dumm09}

With the metal-insulator transition prominently at 80\,K, we do not observe a gradual increase in resistivity of \hgbr{} (Fig.\ \ref{fig:dc}) as known from the dimerized Mott insulators $\kappa$-(BEDT-TTF)$_2$$X$ with Cu-ions in the anion layers, such as \cucl{} or \cucn{}.\cite{Sedlmeier12} This leads us to conclude that the $T=80$\,K transition is not a regular Mott transition. This is supported by the analysis of our optical data and structural considerations.
The rapid change in resistivity must therefore have other reasons, most likely some subtle magnetic ordering. In this regard it is of interest to recall ESR investigations by Yudanova {\it et al.}, who measured hydrogenated and deuterated \hgbr{} and \hgcl{} down to low temperature.\cite{Yudanova94}
After subtracting a Curie contribution, the spin susceptibility vanishes at the metal-insulator transition, following $\chi(T)=C/T \exp\{-\Delta/T\}$, with $\Delta\approx 500$\,K which agrees nicely with the charge gap extracted from our optical and dc transport data. They conclude a first-order structural phase transition with a considerable localization of the electrons on the BEDT-TTF molecules.
Recent low-temperature x-ray diffraction studies, however, did not reveal any evidence of a unit cell doubling or other structural changes in \hgbr{}. Thus we cannot explain the change in transport and magnetic properties by a change in the crystal structure. Comprehensive investigations of magnetic properties are on its way in order to clarify the magnetic ground state of \hgbr{} as well as the Cl-analogue. Additionally, conventional density-wave instabilities are excluded by the absence of structural changes as seen by x-ray and infrared vibrations. Future low-temperature x-ray experiments are needed to explore if subtler structural effects play a role in the fluctuating charge present in this compound.

\section{Conclusions and Outlook}
From our comprehensive transport, dielectric and optical investigations of the dimerized organic charge transfer salt \hgbr, we can rule out that the metal-insulator transition at $T=80$\,K is due to a charge ordering. No pronounced structural change takes place as well. We can estimate the effective correlations to be weaker compared to the $\kappa$-(BEDT-TTF) salts containing Cu-based anions. The system is also more metallic than the sister compound \hgcl. Although electronic correlations do play a decisive role, the phase below the 80\,K transition cannot be explained as a simple Mott insulator. We suggest that magnetic fluctuations might play an important role in supporting charge fluctuations revealed by our infrared vibrational spectra.

\acknowledgments
We do acknowledge valuable discussion with N.\ Drichko and A.\ Pustogow. The project was supported by the Deutsche Forschungsgemeinschaft (DFG) by DR228/39-1 and DR228/52-1, the Deutsche Akademischer Austauschdienst (DAAD), as well as the Croatian Science Foundation project IP-2013-11-1011.

\end{document}